\documentclass[preprint,onecolumn,nofootinbib]{revtex4}
\pdfoutput=1
\usepackage[colorlinks=true,linkcolor=blue,urlcolor=blue,filecolor=black,citecolor=red,pdfstartview=FitV,pdftitle={},pdfsubject={},pdfkeywords={},pdfpagemode=None,bookmarksopen=true]{hyperref}
\usepackage{graphicx}
\usepackage{amsmath}
\usepackage{amsfonts}
\usepackage{amssymb,ulem}
\usepackage{color}%
\usepackage{dcolumn}
\newcommand{\f}{\begin{equation}}
\newcommand{\ff}{\end{equation}}
\newcommand{\fa}{\begin{eqnarray}}
\newcommand{\ffa}{\end{eqnarray}}

\begin{document}
\title{Lifshitz black branes and DC transport coefficients in massive Einstein-Maxwell-dilaton gravity}
\author{Xiao-Mei Kuang$^{1,2}$}
\email{xmeikuang@gmail.com}
\author{Eleftherios Papantonopoulos$^{3}$}
\email{lpapa@central.ntua.gr}
\author{Jian-Pin Wu$^{4}$}
\email{jianpinwu@mail.bnu.edu.cn}
\author{Zhenhua Zhou$^{5}$}
\email{dtplanck@163.com}
\affiliation{\small{$^1$~Center for Gravitation and Cosmology, College of Physical Science and Technology, Yangzhou University, Yangzhou 225009, China}}
\affiliation{\small{$^2$ Instituto de F\'isica, Pontificia Universidad Cat\'olica de Valpara\'iso, Casilla 4059, Valpara\'iso, Chile}}
\affiliation{\small{$^3$ Physics Division, National Technical University of Athens, 15780 Zografou Campus, Athens, Greece}}
\affiliation{\small{$^4$~Institute of Gravitation and Cosmology, Department of
Physics, School of Mathematics and Physics, Bohai University, Jinzhou 121013, China}}
\affiliation{\small{$^5$~School of Physics and Electronic Information, Yunnan Normal University, Kunming, 650500, China}}
\begin{abstract}

We construct  analytical Lifshitz massive black brane solutions in massive Einstein-Maxwell-dilaton gravity theory.
We also study the thermodynamics of these black brane solutions and obtain the thermodynamical stability conditions.
On the dual nonrelativistic boundary field theory with Lifshitz symmetry, we analytically compute the DC transport coefficients,
including the electric conductivity, thermoelectric conductivity, and thermal conductivity.
The novel property of our model is that the massive term supports the
Lifshitz black brane solutions with $z\neq 1$ in such a way that the DC transport coefficients in the dual field theory are finite.
We also find that the Wiedemann-Franz law in this dual boundary field theory is violated,
which indicates that it may involve strong interactions.

  \end{abstract}
\maketitle
\section{Introduction}
In the study of the application of gauge-gravity duality \cite{Maldacena,Gubser,Witten} to condensed matter physics, the introduction of momentum relaxation plays an important role in the development of this research field 
because systems with momentum dissipation are more realistic. Many proposals have  been put forward in order to describe a dual field theory with
momentum relaxation contributing to finite conductivities. They can be classified into two categories.

The first category introduces a spatial-dependent source, so that the Ward identity is broken, which means that momentum is not conserved in the boundary theory.
An obvious way to do this is to introduce a periodic scalar source or chemical potential \cite{Horowitz:2012ky,Horowitz:2012gs,Ling:2013nxa}, called the scalar lattice or ionic lattice structure.
This involves solving partial differential equations (PDEs) in the bulk.
Another important way is to break the translation invariance but hold the homogeneity of the system,
which involves solving only ordinary differential equations (ODEs) in the bulk.
Outstanding examples of this include holographic Q-lattices breaking the translational invariance via the global phase of the complex scalar field \cite{Donos:2013eha,Donos:2014uba,Ling:2015epa,Ling:2015exa},
holographic helical lattices, where the translational invariance is broken in one space direction but holds in the other two space directions, possessing the non-Abelian Bianchi VII$_0$ symmetry \cite{Donos:2012js},
and holographic axion models, for which a pair of spatial-dependent scalar fields are introduced to source the breaking of Ward identity \cite{Andrade:2013gsa,Kim:2014bza,Cheng:2014tya,Ge:2014aza,Andrade:2016tbr,Kuang:2017cgt}.
In addition, by turning on a higher-derivative interaction term between the $U(1)$ gauge field and the scalar field, a spatially dependent
profile of the scalar field is generated spontaneously which leads to the breaking of the Ward identity \cite{Kuang:2013oqa,Alsup:2013kda}. 
In a more general setup, gravitational models were discussed in which on the boundary spontaneous modulation of the electronic charge and spin density is generated by charge and spin density
waves \cite{Aperis:2010cd,Donos:2013gda,Ling:2014saa}.

The second category is to break the diffeomorphism invariance, without additional fields, which leads to the breaking of the conservation of the stress-energy tensor.
Modified gravity theory which breaks diffeomorphism invariance is called massive gravity.  
Vegh first proposed to study the holographic momentum dissipation in this kind of theory \cite{Vegh:2013sk}, which inspired remarkable progress in holographic studies with momentum relaxation in massive gravity \cite{Davison:2013jba,Blake:2013owa,Blake:2013bqa,Amoretti:2014mma,Zhou:2015dha,Mozaffara:2016iwm,Baggioli:2014roa,Alberte:2015isw,Baggioli:2016oqk,Zeng:2014uoa,Amoretti:2014zha,Amoretti:2015gna,Fang:2016wqe}.

On the other hand, in order to holographically study the nonperturbative dynamics of nonrelativistic systems with anisotropic scale invariance $t\to \lambda^z t,~x\to \lambda x~\mathrm{with}~ z\neq 1$, the AdS/CFT duality was first generated in \cite{Kachru:2008yh} to describe a boundary theory with  dynamical critical exponents which shows the dynamical scaling.  The background spacetime to study the nonrelativistic  quantum field theory with dynamical (Lifshitz) scaling, is given by the Lifshitz metric
\begin{eqnarray}\label{eq-LifshitzMetric}
ds^2=-r^{-2z}dt^2+\frac{dr^2}{r^2}+r^{-2}dx_i^2,
\end{eqnarray}
where the spacial index $i$ runs from 1 to $D-2$. The geometry \eqref{eq-LifshitzMetric} recovers anti-de Sitter(AdS) spacetime when $z=1$ and the boundary is at $r\to 0$.

In holographic fashion, it would be interesting to find the bulk sector to dually describe the momentum relaxation of a
nonrelativistic boundary theory, which means we need to add ingredients into the bulk theory to introduce Lifshitz scaling and momentum dissipation. 
To this end, we will construct Lifshitz black brane solutions in the massive Einstein-Maxwell-dilaton (EMD) gravity theory proposed in \cite{Zhou:2015dha}. It is notable that in \cite{Zhou:2015dha}, the authors have studied the dyonic Reissner-Nordstr$\ddot{o}$m AdS(RN-AdS) black brane as well as the hyperscaling violation AdS black brane and then the DC electric and Hall effect in the dual field theory.

In this paper, we are interested in the analytical Lifshitz black brane solutions in the massive EMD gravity theory.
We will focus on two different models distinguished by different couplings in the action. In both cases, we find the analytical black brane solutions by solving the equations of motion, and briefly analyze their thermodynamics. We also analytically compute the DC transport coefficients, including electric, thermoelectric and thermal conductivities in the dual boundary theory. We note that the gravitational description of nonrelativistic theory with momentum relaxation has also been studied in \cite{Andrade:2016tbr,Blake:2015ina,Amoretti:2016cad,Ge:2016lyn,Cremonini:2016avj}, where the authors introduced additional linear axion fields in the bulk to break the translation symmetry on the boundary, belonging to the first category involving momentum relaxation.

The structure of this paper is as follows. We review the holographic massive EMD theory in Sec. \ref{sec:action}. 
Then we construct the Lifshitz  black brane solutions in this framework and study their thermodynamics. 
We continue to explore the DC transport coefficients in the dual theories of two black branes in Sec. \ref{sec:conductivity}. 
The last section is our conclusion and discussion.

\section{Holographic massive Einstein-Maxwell-dilaton theory} \label{sec:action}

We start with a holographic massive EMD action \cite{Zhou:2015dha},
\begin{eqnarray}
S=\int d^4x \sqrt{-g}\left(R-\frac{Z(\phi)}{4}F^2-\frac{1}{2}(\partial\phi)^2+V(\phi)+\beta(\phi)([\mathcal{K}]^2-[\mathcal{K}^2])\right)
\label{ActionAv1}
\,,
\end{eqnarray}
where $\phi$ is the dilaton field with potential $V(\phi)$.
$Z(\phi)$ and $\beta(\phi)$, where all are a function of $\phi$,
are the coupling functions of the gauge field and massive term, respectively.
In the absence of the massive term, this action results from an effective low-energy heterotic string theory after
a conformal transformation of the metric. This transformation introduces a coupling of the dilaton field with the gauge field, and the
dilaton electrically or magnetically charged black holes are known \cite{Garfinkle:1990qj}.
In this work we consider the two massive terms $[\mathcal{K}]^2:=(\mathcal{K}^\mu_{~\mu})^2$ and $[\mathcal{K}^2]:=(\mathcal{K}^2)^\mu_{~\mu}$
with $\mathcal{K}$ satisfying $\mathcal{K}^\mu_\alpha\mathcal{K}^\alpha_\nu=g^{\mu\alpha}f_{\alpha\nu}$
and $(\mathcal{K}^2)_{\mu\nu}:=\mathcal{K}_{\mu\alpha}\mathcal{K}^{\alpha}_{~\nu}$.
$f_{\mu\nu}$ is the reference metric and we are interested in the special case with $f_{\mu\nu}=diag(0,0,1,1)$,
which breaks the symmetries associated with reparametrizations of the spatial $x,y$ coordinates in the bulk.
Correspondingly, it leads to the dissipation of momentum in the dual boundary field theory \cite{Vegh:2013sk}.

Applying the variational approach to the action (\ref{ActionAv1}),
one can derive the Einstein equation, Maxwell equation, and scalar equation as follows:
\begin{eqnarray}
&&
\label{MEQ}
\nabla_{\mu}(Z(\phi)F^{\mu\nu})=0
\,,
\
\\
&&
\label{SEQ}
\Box\phi-\frac{1}{4}\frac{\partial Z}{\partial\phi}F^2+\frac{\partial V}{\partial\phi}+\frac{\partial\beta}{\partial\phi}([\mathcal{K}]^2-[\mathcal{K}^2])=0
\,,
\
\\
&&
\label{EEQ}
R_{\mu\nu}-\frac{1}{2}g_{\mu\nu}\left(R-\frac{Z}{4}F^2-\frac{1}{2}(\partial\phi)^2+V\right)
-\frac{1}{2}\partial_{\mu}\phi\partial_{\nu}\phi-\frac{1}{2}ZF_{\mu\rho}F_{\nu}^{\ \rho}
+X_{\mu\nu}
=0
\,,
\end{eqnarray}
where
\begin{eqnarray}
\label{Xmunu}
X_{\mu\nu}=-\beta(\phi)\left((\mathcal{K}^2)_{\mu\nu}-[\mathcal{K}]\mathcal{K}_{\mu\nu}+\frac{1}{2}g_{\mu\nu}([\mathcal{K}]^2-[\mathcal{K}^2])\right)
\,.
\end{eqnarray}

For some specific forms of $V(\phi), Z(\phi)$, and $\beta(\phi)$, we can obtain the analytical black brane solutions.
In this work, we are interested in the Lifshitz black brane solution. So we shall take the following ansatz to solve the above equations of motion,\begin{eqnarray}
\label{Av1}
&&ds^2=\frac{-f(r)}{r^{2z}}dt^2+\frac{dr^2}{f(r)r^2}+r^{-2}(dx^2+dy^2)\,,\\
\label{Av2}
&&A=A_t(r)dt\,,~~~~~~\phi=\phi(r)
\,,
\end{eqnarray}
where $z$ is the Lifshitz dynamical exponent.

In this paper, we shall focus on two kinds of models:
\begin{itemize}
  \item \textbf{Model I}: The potential and the coupling functions have the exponential forms as
  \begin{alignat}{1}\label{exponent1}
Z(\phi)=Z_0 r^\lambda\,,~~~~~
\beta(\phi)=\beta_0r^{\sigma}\,,~~~~~
V(\phi)=V_0 r^{\gamma}\,,
\end{alignat}
where $(\lambda, Z_0, \sigma, \beta_0,\gamma, V_0)$ are the parameters in this model.
  \item \textbf{Model II}: The forms of the potential and the coupling functions are
  \begin{alignat}{1}\label{exponent2}
Z(\phi)=Z_0 e^{(-2z+2)\phi/\alpha}\,,~~~~~
\beta(\phi)=\beta_0e^{-2\phi/\alpha}+\beta_1\,,~~~~~
V(\phi)=V_0+V_1e^{2\phi/\alpha}\,,
\end{alignat}
where $(Z_0,~\alpha,~\beta_0,~\beta_1,~V_0,~V_1)$ are the parameters of the model.
\end{itemize}

Before proceeding, we would like to present some comments.
First, when we take $\beta_1=0$, the model II reduces to model I as will be show later.
From this point of view, the potential and coupling functions of model II are more general than those of model I,
and so we will mainly present how to obtain the black hole solution and its thermodynamics in model II.
Then we briefly discuss the results in model I.
However, there are two important differences between these two models.
One is that $z=2$ is forbidden in the black brane solution of model II, but it can be obtained in model I.
The other comment is that when $z=1$, the black brane solution of model I reduces to the RN-AdS one,
while model II reduces to that of standard massive gravity
proposed in \cite{Vegh:2013sk} with $\alpha=0$, $\beta=\beta_1$, and $F=m^2=1$.
These two different points can be clearly seen in what follows.
Finally, we emphasize that the models we propose here are different from those for the dyonic hyperscaling violation AdS black brane in \cite{Zhou:2015dha}.
We believe that more interesting models in EMD gravity theory can be analytically solved by adding higher-curvature terms and various couplings of Maxwell fields; see, for examples, \cite{Ghodrati:2014spa,Hendi:2015xya,Zangeneh:2016cbv,Dehyadegari:2017fqo} and therein for the holographic applications.

\subsubsection{The black brane solution}
In model II with \eqref{exponent2}, the Einstein equation reduction $E_t^t-E_r^r$  gives
\begin{equation}\label{Eq-phi2}
\phi^\prime=\alpha/r\,,~~with~~\alpha\equiv2\sqrt{z-1}~,
\end{equation}
with which, the potential and coupling functions  \eqref{exponent2} can be rewritten as
\begin{alignat}{1}
Z(\phi)=Z_0 r^{-2z+2}\,,~~~~~
\beta(\phi)=\beta_0r^{-2}+\beta_1\,,~~~~~
V(\phi)=V_0+V_1r^2\,.
\end{alignat}
Meanwhile, we  solve the Maxwell equation and obtain
\begin{equation}\label{Eq-at2}
A_t(r)=\mu-\frac{Q}{zZ_0}r^{z}~,
\end{equation}
where the constants $\mu$ and $Q$ are identified as  the chemical potential and the charge of the dual boundary theory, respectively.

With the determined fields  (\ref{Eq-phi2}) and (\ref{Eq-at2}), we solve $f(r)$  from the reduced Einstein equations as
\begin{equation}
f(r)=\frac{Q^2r^{2z+2}}{4Z_0z}+\frac{V_0+2\beta_0}{2(z+2)}
+\frac{V_1+2\beta_1}{2z}-Mr^{z+2}\,,
\end{equation}
where $M$ is an integration constant. Then, the equation of motion for the dilaton (\ref{SEQ}) becomes
\begin{equation}\label{im3}
V_0(1-z)-2z\beta_0+r^2((2-z)V_1-2(z-1)\beta_1)=0~.
\end{equation}
Similarly, in order to obtain an asymptotic Lifshitz solution, we find that the parameters should satisfy the
following relations:
\begin{equation}\label{Arelation}
\beta_0=-(z-1)(z+2)\,,~~V_0=2z(z+2)\,,~~V_1=\frac{2(z-1)}{2-z}\beta_1\,.
\end{equation}
Consequently, our model here admits  the Lifshitz black brane solution as follows:
\begin{alignat}{1}\label{solution2}
&f(r)=1-Mr^{z+2}+\frac{Q^2r^{2z+2}}{4Z_0z}+\frac{\beta_1r^{2}}{z(2-z)}\,,\nonumber\\
&A(r)=\mu-\frac{Q}{zZ_0}r^{z}\,,\nonumber\\
&\phi^\prime=\alpha/r\,,~~~with~~\alpha=2\sqrt{z-1}
\,.
\end{alignat}
Since $f(r_h)=0$,  $M$ can be determined as $M=\frac{1}{r_h^{z+2}}+\frac{Q^2 r_h^z}{4zZ_0}-\frac{\beta_1}{z(z-2)r_h^z}$. The temperature of the black brane is calculated as
\begin{equation}\label{T}
T=\frac{1}{4\pi r_h^{z}}\left(z+2-\frac{Q^2r_h^{2+2z}}{4Z_0}+\frac{\beta_1 r_h^2}{2-z}\right)~.
\end{equation}

Note that the Lifshitz exponent $z=2$ is not allowed in the solution \eqref{solution2} of model II
because the blackness function $f(r)$, the mass, and the temperature of the black brane are divergent in this case.
We can, however, require $\beta_1=0$ for $z=2$ in model II.
Immediately, we
note that model II with $\beta_1=0$ recovers model I \eqref{exponent1} with
\begin{alignat}{1}
\lambda=-2z+2,~~~~\sigma=-2~,~~~~\gamma=0~,~~~~\beta_0=-(z-1)(z+2)~,~~~~V_0=2z(z+2)~.
\label{Arelation1}
\end{alignat}

When $z=1$, model II is just the standard massive gravity proposed in \cite{Vegh:2013sk} and Eq. \eqref{solution2}
goes back to the one with $\alpha=0$, $\beta=\beta_1$, and $F=m^2=1$.
For model I, $z=1$ gives $\beta_0=0$. It means that when $z=1$, model I reduces to Einstein-Hilbert gravity and the solution is nothing but the RN-AdS black brane.

\subsubsection{The black brane thermodynamics}
Furthermore, we analyze the thermodynamics of the black brane solution of mode II.
For simplification, we can set $Z_0=1$ without loss of generalization because in both models \eqref{exponent1} and  \eqref{exponent2}, the constant $Z_0$ in action \eqref{ActionAv1} can be absorbed into the gauge field, as does the charge $Q$ or the chemical potential $\mu$. To study the thermodynamical quantities, we write down the Euclidian on-shell action\footnote{The Euclidian on-shell action is obtained via an imaginary time replacement  $t\rightarrow it$ and its period is the inverse of the temperature 1/T.}
\begin{eqnarray}
S_{onshell}
&=&\int d^4x \left(\frac{Q^2 r^{z-1}}{2}-\frac{2z(z+2)}{r^{z+3}}+\frac{2(z-1)\beta_1}{(z-2)r^{z+1}}\right)
\nonumber
\\
&=&\frac{V_2}{T}\left(\frac{Q^2 r^{z}}{2z}+\frac{2z}{r^{z+2}}+\frac{2(1-z)\beta_1}{z(z-2)r^z}\right)\Big|^{r_h}_0
\,,
\end{eqnarray}
where $V_2=\int dxdy$.  In order to cancel the divergence in the on-shell action, we should add the boundary term \cite{Balasubramanian:1999re,Skenderis:2002wp,Tarrio:2012xx,Mozaffar:2012bp}
\begin{alignat}{1}
S_{bdy}=S_{GH}+S_{CT}=-\int_{r\rightarrow0} dx^3\sqrt{-\gamma}(-2K+4)~,
\end{alignat}
where $S_{GH}$ and $S_{CT}$ are Gibbs-Hawking term and counter term, respectively, and $\gamma$ is the determinant of the boundary induced metric $\gamma_{ab}$.  
Then, we deduce the renormalized on-shell action:
\begin{eqnarray}
S^{RN}_{onshell}\equiv S_{onshell}+S_{bdy}=\frac{V_2}{T}\left(\frac{(2-z)Q^2 r_h^{z}}{4z}+\frac{z}{r_h^{z+2}}-\frac{\beta_1}{zr_h^z}\right).
\end{eqnarray}

So far, we are ready to calculate the thermodynamical quantities. The free energy of the system is
\begin{eqnarray}
F\equiv-TS^{RN}_{onshell}=-V_2\left(\frac{(2-z)z^2\mu^2}{4zr_h^{z}}+\frac{z}{r_h^{z+2}}-\frac{\beta_1}{zr_h^z}\right)~,
\end{eqnarray}
where we have used $Q=\mu z /r_h^z$ because of the regular condition $A_t(r_h)=0$. Then, we have the pressure
\begin{eqnarray}
P\equiv-\left(\frac{\partial F}{\partial V_2}\right)_{\mu,r_h}=\frac{(2-z)z^2\mu^2}{4zr_h^{z}}+\frac{z}{r_h^{z+2}}-\frac{\beta_1}{zr_h^z}\,
\end{eqnarray}
and the entropy
\begin{eqnarray}
S\equiv \left(\frac{\partial F}{\partial T}\right)_{\mu,V_2}=\frac{\left(\partial F/\partial r_h\right)_{\mu,V_2}}{\left(\partial T/\partial r_h\right)_{\mu,V_2}}=\frac{4\pi V_2 }{r_h^{2}}\,.
\end{eqnarray}
According to the first law of thermodynamics, i.e., $\epsilon V_2 + P V_2= ST + \mu Q$, we can deduce the energy density:
\begin{eqnarray}
\epsilon=\frac{z\mu^2}{2r_h^{z}}+\frac{2}{r_h^{z+2}}+\frac{2\beta_1}{z(2-z)r_h^{z}}
~.
\end{eqnarray}
Then, the specific heat is
\begin{eqnarray}
c_V\equiv \left(\frac{\partial \epsilon}{\partial T}\right)_{\mu}=4\pi \frac{\frac{z^2\mu^2}{2}+2(z+2)r_h^{-2}+\frac{2\beta_1}{2-z}}{z(z+2)+\frac{z^2(2-z)\mu^2r_h^2}{4}+\beta_1 r_h^2}
\,.
\end{eqnarray}
Thus, the condition of the thermodynamical stability of the black brane is
\begin{equation}\label{eq:condtion1}
z(z+2)+\left(\frac{z^2(2-z)\mu^2}{4}+\beta_1\right)r_h^2>0,
\end{equation}
which implies positive specific heat.

We note that all the thermodynamic properties for the black brane in model I can be reduced by putting $\beta_1=0$ in all the thermodynamic quantities above.

\section{The DC transport coefficients}\label{sec:conductivity}

We will explore the DC transport coefficients, including the DC electric $\sigma$, thermoelectric $\alpha$, and thermal conductivities $\bar{\kappa}$, of the boundary field theory dual to the black brane solution of model II, while the DC transport coefficients of model I can be extracted by taking $\beta_1=0$.
The method we adopted here is the one developed in \cite{Donos:2014cya} which is a powerful tool to calculate the DC transport coefficients.
To this end, we turn on the following consistent perturbations at the linearized level, which source the electric and heat currents
\begin{equation}\label{eq-perturbation}
A_x:= a_t t+a_x\,,\qquad\delta g_{tx}:= h_{tx}t+H_{tx}\,,\qquad \delta g_{rx}:=H_{rx}/r^2\,,
\end{equation}
where $a_t$, $a_x$, $h_{tx}$, $H_{tx}$, and $H_{rx}$ are all functions of $r$.
Then, the Maxwell equation for the perturbation are
\begin{eqnarray}\label{eq-per-ax}
&& f r^2 (a_x''+a_t''t)+ \left[r^2 f'-3 f r (z-1)\right](a_x' +a_t't) \nonumber \\
  &-&Q r^{3 z}\left[r (H_{\text{tx}}'
+ h_{\text{tx}}'t)+2  (H_{\text{tx}}+h_{\text{tx}}t)\right]=0~.
\end{eqnarray}
The nonvanishing linear Einstein equations for perturbations are the $rx$ component,
\begin{eqnarray}\label{eq-per-hrx}
H_{\text{rx}}+\frac{r^{z+1} \left(Q r^2 a_t-r^z \left(r h_{\text{tx}}'+2
   h_{\text{tx}}\right)\right)}{2 f \left(z^2+z-2-\beta _1r^2\right)}=0~,
\end{eqnarray}
and the $tx$ one
\begin{eqnarray}\label{eq-per-htx}
&&f r^{1-z}\left(H_{\text{tx}}''+ h_{\text{tx}}''t\right)+f r^{-z} (z+1) \left(H_{\text{tx}}'+ h_{\text{tx}}'t\right)-f Q r^{2-2z} \left(a_x'+ a_t't\right) \\
&&+\left(2z+4-2(z^2+z+2)f+3r(z+1)f'-r^2f''+\frac{Q^2r^{2(z+1)}}{2} -\frac{2 \beta _1 r^2}{z-2}\right)\frac{H_{\text{tx}}+ h_{\text{tx}}t}{r^{z+1}}=0~.\nonumber
\end{eqnarray}

The key point of the method in \cite{Donos:2014cya}
is to find the conserved electric current $J(r)$ and the conserved heat current $\mathcal{Q}(r)$ such that we can evaluate them at the horizon.
In our present model, we find that Eqs. \eqref{eq-per-ax} and \eqref{eq-per-htx} can be written as the integral forms $\frac{d J(r)}{d r}=0$ and $\frac{d \mathcal{Q}(r)}{d r}=0$ with
\begin{eqnarray}
&&
\label{eq-Jx}
J(r)=\frac{Q r^{3 z}(H_{\text{tx}}+ h_{\text{tx}}t)-f r( a_x'+ a_t't)}{r^{3 z-2}}\,,\ \\
&&
\label{eq-Qx}
\mathcal{Q}(r)=r^{1-z}\left[f (H_{tx}'+h_{tx}'t)-\left(f'-\frac{2z}{r}f  \right)(H_{tx}+h_{tx}t)  \right]-A_tJ\,.
\end{eqnarray}
We then further assume $h_{tx}=-\xi r^{-2z} f(r)$ and $a_t=-E+\xi A_t(r)$, where the constants $E$ and $\xi$  parametrize the sources for the electric current and heat current, respectively \cite{Donos:2014cya}. Subsequently, we obtain the time-independent terms of the conserved currents:
\begin{eqnarray}\label{eq-NewCurrent}
\bar{J}&=&\frac{Q r^{3 z}H_{\text{tx}}-f r a_x'}{r^{3 z-2}}~,\\
\bar{\mathcal{Q}}&=&r^{1-z}\left[f H_{tx}'-\left(f'-\frac{2z}{r}f  \right)H_{tx} \right]-A_t \bar{J}~.
\end{eqnarray}

According to the AdS/CFT dictionary, the DC conductivities are determined by the conserved currents in the asymptotical boundary accompanying the regularity conditions at the horizon for the perturbations.
As mentioned above, since $\bar{J}$ and $\bar{\mathcal{Q}}$ are conserved quantities in the bulk and are $r$ independent,
we shall evaluate them at the horizon.
In addition, to impose the regularity conditions at the horizon for perturbations,
it is convenient to transfer to Eddington-Finklestein coordinates $(v,r)$ with $v=t+\int \frac{r^{z-1}}{f}dr$.
Thus, the regular condition for the gauge field denotes that at the future horizon $A_x\propto v $, and recalling $A_x$
in \eqref{eq-perturbation} and $a_t=-E+\xi A_t(r)$, we obtain
\begin{eqnarray}\label{eq-fieldsH1}
a_x=-E \int \frac{r^{z-1}}{f}dr\,
\end{eqnarray}
near the horizon. Besides, in order to avoid the singularity in the metric
\begin{equation}
2H_{tx}dvdx-\frac{2r^{z-1}}{f}H_{tx}drdx+\frac{2H_{rx}}{r^2}drdx,
\end{equation}
 we have to impose the following relation between the perturbations near the horizon
\begin{eqnarray}\label{eq-fieldsH2}
H_{tx}=\frac{f}{r^{z+1}}H_{rx}~,
\end{eqnarray}
where $H_{rx}$ satisfies Eq. \eqref{eq-per-hrx}. Recalling that, near the horizon, we have $f|_{r\to r_h}\sim 4\pi T r_h^{z-1} (r-r_h) $ and considering \eqref{eq-fieldsH1} and  \eqref{eq-fieldsH2}, we obtain the currents evaluated at the horizon:
\begin{eqnarray}\label{eq-NewCurrentH}
\bar{J}&=&\frac{E(Q^2r_h^{2z+2}+2(z^2+z-2-r_h^2\beta_1))-4\pi T Q\xi r_h^{2z}}{2r_h^{2z-2}(z^2+z-2-r_h^2\beta_1)}~,\\
\bar{\mathcal{Q}}&=&-\frac{2\pi T E Q r_h^2-8\pi^2T^2\xi}{z^2+z-2-r_h^2\beta_1}~.
\end{eqnarray}

Subsequently, we are ready to compute the DC transport coefficients as addressed in \cite{Donos:2014cya}. The DC electric conductivity is
 \begin{equation}\label{eq-sigma}
 \sigma_{DC} =\frac{\partial \bar{J}}{\partial E}=r_{h}^{2-2z}-\frac{Q^2 r_h^4}{4-2z-2z^2+2r_h^2 \beta_1}~.
 \end{equation}
The two thermoelectric conductivities are
 \begin{eqnarray}\label{eq-alpha}
 \alpha_{DC}& =&\frac{1}{T}\frac{\partial \bar{J}}{\partial \xi}=\frac{2\pi Q r_h^2}{r_h^2 \beta_1-(z^2+z-2)}~,\\
 \bar{\alpha}_{DC}& =&\frac{1}{T}\frac{\partial \bar{\mathcal{Q}}}{\partial E}=\frac{2\pi Q r_h^2}{r_h^2 \beta_1-(z^2+z-2)}= \alpha_{DC}~.
 \end{eqnarray}
And the thermal conductivity is
 \begin{eqnarray}\label{eq-kappa}
\bar{\kappa}_{DC} =\frac{1}{T}\frac{\partial \bar{\mathcal{Q}}}{\partial \xi}=\frac{8\pi^2 T}{(z^2+z-2)-r_h^2 \beta_1}~.
 \end{eqnarray}
Note that the DC electric conductivity \eqref{eq-sigma} with $z=1$ recovers the result obtained in \cite{Blake:2013bqa}
for massive Einstein gravity.
All the DC transport coefficients are affected by the Lifshitz exponent and the massive parameters.
Explicitly, larger $z$ suppresses both $\sigma_{DC}$ and $\kappa_{DC}/T$ while large $\beta_1$ enhances them; however, their affects on $ \alpha_{DC}(\bar{\alpha}_{DC})$ behave in an opposite way.

It is easy to check that the conductivities for model I can be obtained when we let $\beta_1=0$ for model II.
Here, we would like to point out the properties of the DC transport coefficients in these two models:
\begin{itemize}
  \item All the conductivities become divergent when the condition $2-z-z^2+r_h^2 \beta_1=0$ is satisfied. The effective mass of the graviton $\beta(\phi)=\beta_0 r^{-2}+\beta_1$ with $\beta_0=-(z-1)(z+2)$ in \eqref{Arelation} vanishes at the horizon when $2-z-z^2+r_h^2 \beta_1=0$, which means that the translational symmetry is recovered in the IR limit.

\item When $z=1$,
the DC conductivities are divergent for $\beta_1=0$, while they are finite for $\beta_1\neq 0$. In this case, $\beta_0$ vanishes [see Eq.\eqref{Arelation}], and so the massive term, which is responsible for the momentum dissipation, only depends on the parameter $\beta_1$.

  \item When the system has the Lifshitz symmetry, i.e., $z\neq 1$,
  the DC conductivities are finite even though $\beta_1=0$. This is because $z\neq 1$ implies $\beta_0\neq 0$ in \eqref{Arelation}, so that the massive term still survives.\footnote{The case for $z=2$ in model II is an exception, in which both $\beta_0$ and $\beta_1$ vanish.} This property in our models is novel because our solution with $z\neq 1$ presents holographic momentum relaxation; in usual Lifshitz gravity, it cannot be fulfilled.

\end{itemize}

Furthermore, we will briefly check the Wiedemann-Franz (WF) law. It states in \cite{Mahajan:2013cja} that the ratio of the electronic contribution of the thermal conductivity to the electrical conductivity in a conventional metal is proportional to the temperature,\footnote{The law reflects that, for Fermi liquids, the ability of the quasiparticles to transport heat is determined by their ability to transport charge, so that the Lorenz ratio is a constant.} i.e., $L=\frac{\kappa}{\sigma T}=\frac{\pi^2 e^2}{3k_B^2}$ where $\kappa$ is the thermal conductivity at vanishing electric current. In our model, we have
\begin{equation}
\kappa=\bar{\kappa}-\frac{\alpha \bar{\alpha}T}{\sigma}=\frac{16 \pi ^2 T}{Q^2 r_h^{2 z+2}+2
   \left(z^2+z-2\right)-2 \beta _1 r_h^2}~.
\end{equation}
Then, the Lorenz ratios are
\begin{eqnarray}
\bar{L}&=&\frac{\bar{\kappa}}{\sigma T}=\frac{16 \pi ^2 r_h^{2 z-2}}{Q^2 r_h^{2 z+2}+2
 \left(z^2+z-2\right)-2 \beta _1 r_h^2}~,\\
 L&=&\frac{\kappa}{\sigma T}=\frac{32 \pi ^2 r_h^{2 z-2} \left((z^2+z-2)-\beta _1 r_h^2\right)}{\left[Q^2 r_h^{2 z+2}-2 \beta _1 r_h^2+2
   \left(z^2+z-2\right)\right]{}^2}~,
\end{eqnarray}
which are not constant and explicitly depend on the Lifshitz exponent and the massive parameters. Therefore, in our models, the WF law is violated, which indicates that our dual systems may involve strong interactions as discussed in \cite{Donos:2014cya}.

\section{Conclusion and discussion}
In this paper, we constructed analytical Lifshitz black brane solutions in the massive EMD gravity and obtained their thermodynamical stability condition by studying black brane thermodynamics.
We then analytically calculated the DC transport coefficients, including electric, thermoelectric and thermal conductivities in the dual boundary field theory.

Our results are summarized as follows:
\begin{itemize}

\item In model I, when the Lifshitz exponent $z=1$, the model recovers Einstein-Hilbert gravity and the DC transport coefficients are divergent as expected.
While when $z\neq 1$, model I admits a Lifshitz massive black brane solution. The DC conductivities
in the dual boundary field theory are finite because $z\neq 1$ causes the massive term in the action to survive. This observation in our model is novel because the usual Lifshitz black brane with $z\neq 1$, which has no mechanism for momentum relaxation, cannot produce finite conductivities.

\item In model II, when the Lifshitz exponent is $z=1$, the model reproduces the massive gravity proposed in \cite{Vegh:2013sk} with the parameters $\alpha=0$, $\beta=\beta_1$, and $F=m^2=1$. The DC transport coefficients are all finite in both cases with $z=1$ and $z\neq1$ because $\beta_1\neq 0$ corresponds to the nonvanished massive term. Moreover, from the expressions \eqref{eq-sigma}$-$\eqref{eq-kappa}, it is obvious that when $2-z-z^2+r_h^2 \beta_1=0$, the conductivities are divergent because $2-z-z^2+r_h^2 \beta_1=0$ implies that the massive term in the action vanishes at the horizon and the translational symmetry is not broken in the dual IR theory.

\item Our model II with $\beta_1=0$ goes back to model I. This happens because putting the value  $z=2$ in model II, the finite blackness function, mass and temperature of the solution forces $\beta_1=0$ and $V_1=0$, which leads to model I.

\item Finally, we found the Wiedemann-Franz (WF) law is violated in our models by studying the Lorenz ratios. This implies that the dual systems of the models may involve strong interactions.
\end{itemize}

It shall be very interesting to compare the features we found in the boundary field theory with momentum relaxation dual to our gravitational models with the experimental observables in the real materials, such as the condensed matter systems governed by nonrelativistic conformal field theories \cite{Schdev}. In the future, we shall further explore this and mimic the possible materials that possess the properties in our nonrelativistic studies.

The Lifshitz massive black branes we found are new and it is natural to extend them to wider holographic applications.
We believe that more interesting analytical black brane solutions can be found in the EMD gravity.

\begin{acknowledgments}
We would like to thank Peng Liu for the stimulating discussions.
This work is supported by the Natural Science
Foundation of China under Grants No. 11705161, No. 11775036, No. 11747038, and No. 11305018.
X. M. Kuang is also supported by Natural Science Foundation of Jiangsu Province under Grant No. BK20170481 and she is also funded by Chilean FONDECYT Grant No. 3150006.
J. P. Wu is also supported by the Natural Science Foundation of Liaoning Province under Grant No.201602013.
\end{acknowledgments}


\end{document}